\documentclass[aps,prl,twocolumn,floatfix,english,superscriptaddress,showpacs,10pt]{revtex4-2} 
\usepackage{amsmath}
\usepackage{mathtools}
\usepackage{commath}
\usepackage{amssymb}
\usepackage{graphicx}
\usepackage{babel}
\usepackage{cases}
\usepackage[colorlinks=true, citecolor=blue, anchorcolor=green]{hyperref}
\hypersetup{linkcolor=red}
\usepackage{natbib}
\usepackage{dblfloatfix}
\usepackage{xcolor}
\usepackage{braket}
\usepackage{lipsum}
\usepackage{esint}

\makeatletter

\newcommand{\Rmnum}[1]{\expandafter\@slowromancap\romannumeral #1@}

\makeatother

\begin{document}

\title{Evanescent Orbital Pumping by Magnetization Dynamics Free of Spin-Orbit Coupling}

\author{Chengyuan Cai}
\affiliation{School of Physics, Huazhong University of Science and Technology, Wuhan 430074, China}

\author{Hanchen Wang}
\affiliation{Department of Materials, ETH Zurich, Zurich 8093, Switzerland}

\author{Tao Yu}
\email{taoyuphy@hust.edu.cn}
\affiliation{School of Physics, Huazhong University of Science and Technology, Wuhan 430074, China}
\date{\today }

\begin{abstract}
Converting magnetization spin to orbital current often relies on strong spin-orbit interaction that may cause additional angular momentum dissipation. We report that coherent magnetization dynamics in magnetic nanostructures can evanescently pump an orbital current into adjacent semiconductors due to the coupling between its stray electromagnetic field and electron orbitals without relying on spin-orbit coupling. The underlying photonic spin of the electromagnetic field governs the orbital polarization that flows along the gradient of the driven field. Due to the joint effect of the electric and magnetic fields, the orbital Hall current that flows perpendicularly to the gradient of the time-varying field is also generated and does not suffer from the orbital torque. These findings extend the paradigm of orbital pumping to include photonic angular momentum and pave the way for developing low-dissipation orbitronic devices.
\end{abstract}

\maketitle

\textit{Introduction}.---Controlling and harnessing angular momentum in quantum materials is a central theme in spintronics~\cite{manchon2019current,chumak2015magnon,demidov2017magnetization,bauer2012spin,maekawa2023spin,sinova2015spin,bihlmayer2022rashba,ralph2008spin,chirality}. Spin pumping is the generation of spin current driven by coherent magnetization dynamics~\cite{tserkovnyak2005nonlocal,harder2016electrical} that leads to significant advances in controlling magnetization and practical applications of spintronics~\cite{tombros2007electronic,kajiwara2010transmission,yang2008giant,miron2011perpendicular,liu2012spin,caretta2018fast,grimaldi2020single,torrejon2017neuromorphic,krizakova2022spin,sun2022spin,blasing2020magnetic,luo2020current,WenTianLu,Thermal_Magnon_Transistors}. 
While spin degrees of freedom have traditionally occupied the spotlight, recent studies suggest that the orbital degrees of freedom in solids can play an equally crucial role in angular momentum transport. Indeed, in many crystalline materials, such as those with diamond or zinc-blende structures, the valence bands possess rich orbital textures inherited from the underlying lattice symmetries~\cite{Luttinger1955,Luttinger1956,bernevig2005orbitronics,Bhowal1,Bhowal2,Vignale1,Pezo,Culcer}. 
The emerging field of ``orbitronics" aims to exploit the intrinsic orbital character of electronic bands for new device functionalities in systems where conventional spin-orbit coupling (SOC) is weak or absent~\cite{bernevig2005orbitronics,tanaka2008intrinsic,kontani2009giant,go2018intrinsic,han2022orbital,salemi2022firstprinciples,pezo2024theory,choi2023orbital,lyalin2023magneto,sala2023orbital,ding2022observation}.

The orbital pumping, whereby a dynamic perturbation (e.g., a time-dependent magnetization) injects orbital angular momentum (OAM) into an adjacent nonmagnetic medium, has been proposed and recently observed~\cite{go2023orbital,han2025theory,hayashi2023pump,elhamdi2023observation,wang2025BiYIG,huang2024orbital,santos2024exploring,santos2023inverse,santos2024bulk,abrao2025anomalous}. 
The orbital Hall effect and orbital torque~\cite{go2020orbital,go2023longrange, gao2018intrinsic,lee2021orbital,ding2020harnessing,lee2021efficient,sala2022giant,hayashi2023observation,nikolaev2024large,ding2024orbital,zheng2020magnetization,yang2024orbital,mendoza2024efficient,Rothschild} 
have been performed recently in the light nonmagnetic metals~\cite{sala2022giant,Rothschild,choi2023orbital,lyalin2023magneto}, heavy nonmagnetic metals~\cite{sala2022giant},  semiconductor materials (Si, Ge)~\cite{Santos_Ge,Matsumoto}, transition metal dichalcogenides~\cite{cysne2022orbital,cysne2021disentangling,canonico2020two,costa2023connecting,canonico2020orbital}, and antiferromagnetic materials~\cite{abrao2025anomalous}.
It turns out that the SOC cannot be avoided when generating orbital current by magnetization dynamics and orbital torques to the magnetization~\cite{Culcer_review}, which may cause an additional dissipation of angular momentum. 
Indeed, the electron orbitals cannot directly interact with magnetization through exchange, which indicates that the pumping of orbital current by magnetization dynamics relies on the conversion between spin and orbital currents by SOC. On the other hand, the SOC is needed to convert orbital accumulation
into spin accumulation that creates a torque~\cite{go2020orbital,go2023longrange, gao2018intrinsic,lee2021orbital,ding2020harnessing,lee2021efficient,sala2022giant,hayashi2023observation,nikolaev2024large,ding2024orbital,zheng2020magnetization,yang2024orbital,mendoza2024efficient,Rothschild}.

\begin{figure}[htp!]
\centering
\includegraphics[width=0.45\textwidth]{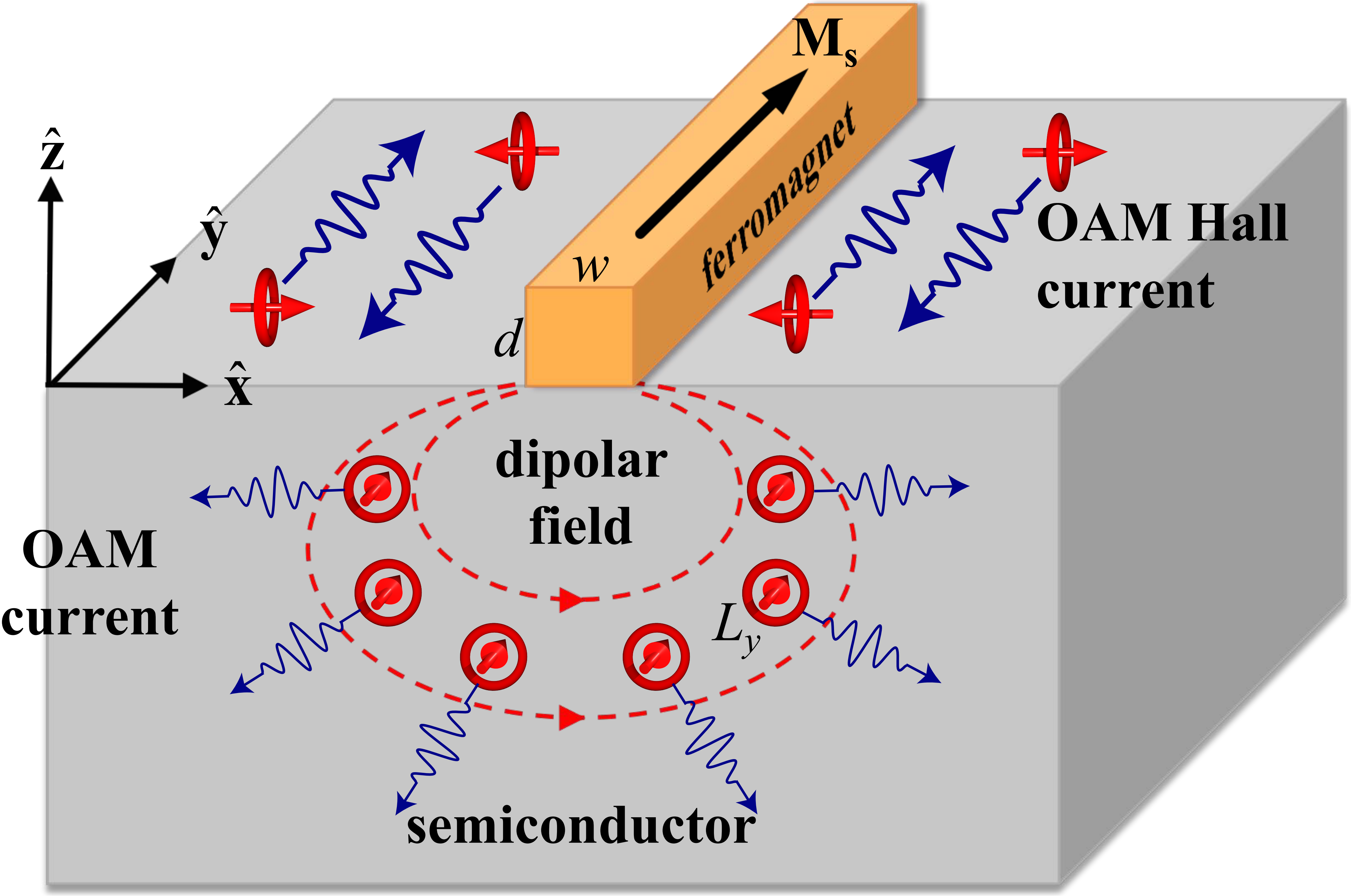}
\caption{Pumping of longitudinal and Hall OAM currents in semiconductors by the AC magnetic field generated by, \textit{e.g.}, the electromagnetic radiation of magnetic nanostructures. The origin of the coordinate is located at the interface below the center of the nanowire. Under a local AC electromagnetic field (red dashed curve) generated by the magnetization precessing in a biased nanowire, a longitudinal OAM current (front blue arrows) with $\hat{\bf y}$-direction orbital polarization (front red arrows) is radiated from the local source in the $x$-$z$ plane. A Hall orbital current (top blue arrows) with $\hat{\bf x}/\hat{\bf z}$-direction orbital polarizations (top red arrows) is pumped along the wire $y$-axis.}
\label{model}
\end{figure}

In this work, we propose a mechanism entirely free of the SOC for directly pumping the orbital current through magnetization dynamics (or a local AC electromagnetic field of an antenna) that is feasible in a hybrid structure with magnetic/metallic nanostructures coupled to semiconductors, as illustrated in Fig.~\ref{model}. 
We go beyond the conventional orbital pumping paradigm by predicting that via the interaction between the time-dependent electromagnetic field and the orbital magnetic moments of electrons, the local dynamic magnetization can effectively transfer its photonic spin to the OAM of electronic states in semiconductors/metals. The injected DC orbital current, polarized along the photonic spin of the AC electromagnetic field, flows along the gradient of the AC field. Such a flow is accompanied by an intrinsic orbital torque to the OAM, arising from the OAM's non-conservation due to the crystal field. Crucially, we find that an orbital Hall current is also pumped, which flows perpendicular to the gradient of the AC field and does not suffer from the orbital torque. Such orbital pumping is evanescent without relying on the contact exchange interaction that may be difficult to achieve between ferromagnets and semiconductors due to Schottky barriers and electronic structure mismatches, providing a promising basis for designing novel orbitronic devices with low power consumption.

\textit{Model}.---We illustrate the principle of the evanescent orbital pumping using the Luttinger model~\cite{Luttinger1955,Luttinger1956}. 
To this end, we take the prototypical semiconductors such as Si~\cite{Santos_Ge,Matsumoto} as examples, which contain three dominant $p$-orbitals at the $\Gamma$ point of valence bands. Our formalism applies to general materials with multi-orbitals at the Fermi surface.
Assuming the SOC is weak~\cite{Luttinger1955,bernevig2005orbitronics} when the resulted split-off energy is smaller that the chemical potential $\mu$, we adopt a spherical approximation, resulting in an effective Hamiltonian of electrons with wave vector 
${\bf k}$~\cite{bernevig2005orbitronics} 
\begin{align}
    {H}_0({\bf k})=A k^2-r({\bf k}\cdot\boldsymbol{\mathcal{I} })^2,
\end{align}
where the coefficients $A$ and ${r}$ are given in the L\"{o}wdin partitioning theory~\cite{Lowdin,Luttinger1956,winkler,GroupTheory} and the OAM matrices   
\begin{align}
    {\cal I}_x&=\left(\begin{array}{ccc}
0 & 0 & 0\\
0 & 0 & -i \\
0 & i & 0
\end{array}\right),~~~{\cal I}_y=\left(\begin{array}{ccc}
0 & 0 & i\\
0 & 0 & 0 \\
-i & 0 & 0
\end{array}\right),\nonumber\\
{\cal I}_z&=\left(\begin{array}{ccc}
0 & -i & 0\\
i & 0 & 0 \\
0 & 0 & 0
\end{array}\right).
\nonumber
\end{align}
The helicity ${\bf e}_ {\bf k}\cdot\boldsymbol{\mathcal{I}}$ along the propagation direction ${\bf e}_{\bf k}=\sin\phi_{\bf k}\cos\theta_{\bf k}\hat{\bf x}+\cos\phi_{\bf k}\hat{\bf y}+\sin\phi_{\bf k}\sin\theta_{\bf k}\hat{\bf z}$ commutes with ${H}_0({\bf k})$ and is thereby conserved.
Among the three valence bands, two bands ``$a$" and ``$b$" are degenerate with dispersions $\varepsilon_{a/b}({\bf k})=(A-r)k^2$ and the third one ``$c$" has a dispersion $\varepsilon_c({\bf k})=Ak^2$. The associated effective masses of heavy and light holes are $m_a^h=m_b^h=\hbar^2/|2(A-r)|$ and $m^h_c=\hbar^2/|2A|$.
The Bloch states of three bands $\psi_{\alpha,{\bf k}}({\bf r})=(1/\sqrt{V})e^{i{\bf k}\cdot{\bf r}}\varphi_{\alpha,{\bf k}}$, in which $V$ is the crystal volume and $\varphi_{a,{\bf k}}=(-\cos\phi_{\bf k}\sin\theta_{\bf k},\sin\phi_{\bf k},-\cos\phi_{\bf k}\cos\theta_{\bf k})^T$, $\varphi_{b,{\bf k}}=(\cos\theta_{\bf k},0,-\sin\theta_{\bf k})^T$, and $\varphi_{c,{\bf k}}=(\sin\phi_{\bf k}\sin\theta_{\bf k},\cos\phi_{\bf k},\sin\phi_{\bf k}\cos\theta_{\bf k})^T$ (see Supplementary Material for the details). 
By the hole operator  $\hat{h}_{\alpha,{\bf k}}$ $ (\hat{h}^\dagger_{\alpha,{\bf k}})$ that annihilates (creates) a hole in the Bloch band $\alpha$ with wave vector ${\bf k}$ and energy $\varepsilon^h_{\alpha,{\bf k}}=(\hbar^2k^2)/(2m^h_\alpha)$, the hole Hamiltonian reads (refer to the Supplementary Material for the construction) 
\begin{align}
    \hat{H}_0=\sum_{\alpha=\{a,b,c\}}\sum_{\bf k}(\varepsilon^h_{\alpha,{\bf k}}-\mu)\hat{h}^\dagger_{\alpha,{\bf k}}\hat{h}_{\alpha,{\bf k}}.
    \label{Luttinger_2D_A=B}
\end{align}

An electromagnetic field $\{{\bf E},{\bf H}\}({\bf r},t)$ that is emitted by, \textit{e.g.}, the magnetization dynamics of magnetic nanostructures~\cite{Baumgaertl,wang2025BiYIG} or an antenna couples to the electron orbitals. To be a source of the orbital current, the time-varying electromagnetic field has to be \textit{local}. By the Kohn-Luttinger transcription
$\hbar{\bf k}\rightarrow\hbar\hat{\bf K}=({\hbar}/{i})\nabla-e{\bf A}$~\cite{Luttinger1955,Luttinger1956,GroupTheory}, the commutator of momentum operators  $[\hat{K}_a,\hat{K}_b]=i\epsilon_{abc}\mu_0eH_c/\hbar$, where $\mu_0$ is the vacuum permeability and $\epsilon_{abc}$ is the antisymmetric tensor, generates an additional Zeeman coupling $\mu_0\mu_B {\bf H}({\bf r},t)\cdot \pmb{\cal L}$ between electron orbitals and external magnetic field (refer to Supplementary Material for derivations)~\cite{Luttinger1956,GroupTheory}, in which the \textit{effective} OAM matrix 
\[
{\cal L}_{\alpha}=(m_0D/\hbar^2){\cal I}_\alpha.
\]
Here, $m_0$ is the mass of free electrons,  $\mu_B=|e|\hbar/(2m_0)$ is the Bohr magneton, and $D$ is another material coefficient governed by the band mixing~\cite{Luttinger1956}, which is detailed in the Supplementary Material. 
With the hole field operator 
$\hat{\Psi}({\bf r})=\sum_{\alpha,{\bf k}}\psi_{\alpha,{\bf k}}({\bf r})\hat{h}^\dagger_{\alpha,-{\bf k}}$, the Zeeman interaction implies the OAM operator 
\begin{align}
    \hat{\bf L}=\int d{\bf r}\hat{\Psi}^\dagger({\bf r})\hbar\boldsymbol{\cal L}\hat{\Psi}({\bf r})=\sum_{\alpha,\beta}\sum_{\bf k}\hbar\tilde{\boldsymbol{\cal L}}^h({\bf k})|_{\alpha\beta}\hat{h}^\dagger_{\alpha,{\bf k}}\hat{h}_{\beta,{\bf k}},
\end{align}
 where $\tilde{\boldsymbol{\cal L}}^h({\bf k})=-(U^\dagger(-{\bf k})\boldsymbol{\cal L}U(-{\bf k}))^T$ is the OAM carried by holes of wave vector ${\bf k}$, governed by the unitary transformation $U({\bf k})=(\varphi_{a,{\bf k}},\varphi_{b,{\bf k}},\varphi_{c,{\bf k}})$. $\tilde{\boldsymbol{\cal L}}^h({\bf k})$ depends strongly on the wave-vector direction, giving rise to an orbital texture in the ${\bf k}$ space, which plays an important role in the orbital Hall effect~\cite{bernevig2005orbitronics,go2018intrinsic}. The OAM  derived from the Kohn-Luttinger transcription in the ${\bf k}\cdot{\bf p}$ Hamiltonian includes both the atomic and itinerant contributions, consistent with the general definition of OAM~\cite{GWang,Pezo,Pezo2023,Culcer,QNiu1999,Culcer_review}. In the Supplementary Material, we compare the Zeeman coupling to the spin and orbital of holes. The Zeeman coupling to hole spin is weaker than the orbitals~\cite{Luttinger1956,Lawaetz}.

The monochromatic electric ${\bf E}({\bf r},t)={\bf E}^{(+)}({\bf r})e^{-i\omega t}+{\bf E}^{(-)}({\bf r})e^{i\omega t}$ and magnetic 
${\bf H}({\bf r},t)={\bf H}^{(+)}({\bf r})e^{-i\omega t}+{\bf H}^{(-)}({\bf r})e^{i\omega t}$ field of frequency $\omega$ drive the orbitals via 
\begin{align}
\hat{H}_{\rm int}=\sum_{\zeta=\pm}\sum_{\alpha,\beta}\sum_{\bf k,k'}e^{-i\zeta\omega t}G^{h,(\zeta)}_{\bf kk'}|_{\alpha\beta}\hat{h}^\dagger_{\alpha,{\bf k}}\hat{h}_{\beta,{\bf k}'},
\label{Zeeman}
\end{align}
in which with 
\begin{align}
\tilde{\boldsymbol{\cal L}}^h_{\bf kk'}\equiv -(U^\dagger(-{\bf k}')\boldsymbol{\cal L}U(-{\bf k}))^T
\end{align}
and 
\begin{align}
\tilde{\boldsymbol{\cal T}}^h_{\bf kk'}&\equiv (U^\dagger(-{\bf k}')(A({\bf k}+{\bf k}')\nonumber\\
&-\sum_{j=\{x,y,z\}}({r}/{2})\{\boldsymbol{\cal I},{\cal I}_j\}(k_j+k_j'))U(-{\bf k}))^T,
\end{align}
the coupling matrix 
\begin{align}
G^{h,(\zeta)}_{\bf kk'} &=(\mu_0\mu_B/V){\bf H}^{(\zeta)}({\bf k-k'})\cdot \tilde{\boldsymbol{\cal L}}^h_{\bf kk'}\nonumber\\
&+(\zeta ie/\hbar\omega V){\bf E}^{(\zeta)}({\bf k-k'})\cdot\tilde{\boldsymbol{\cal T}}^h_{{\bf kk'}}
\end{align}
contains both contributions from electric and magnetic fields. Here,  
${\bf H}^{(\pm)}({\bf k})[{\bf E}^{(\pm)}({\bf k})]=\int d{\bf r} {\bf H}^{(\pm)}({\bf r})[{\bf E}^{(\pm)}({\bf r})]e^{-i{\bf k}\cdot{\bf r}}$ denotes the wave-vector components of the magnetic/electric field integrated over the crystal volume.

With the Bloch states of valence bands $\{a,b,c\}$, the OAM density operator
\begin{align}
    \hat{\bf L}_d({\bf r})&=\hat{\Psi}^\dagger({\bf r})\hbar\boldsymbol{\cal L}\hat{\Psi}({\bf r})\nonumber\\
    &=\frac{1}{V}\sum_{\alpha,\beta}\sum_{\bf k,k'}\hbar\tilde{\boldsymbol{\cal L}}^h_{\bf kk'}|_{\alpha\beta}e^{i({\bf k'-k})\cdot {\bf r}}\hat{h}^\dagger_{\alpha,{\bf k}}\hat{h}_{\beta,{\bf k'}}
\label{OAMD_operator}
\end{align}
defines the OAM current density $ \hat{\bf J}_d({\bf r})$ and OAM torque density $\hat{\bf T}_d({\bf r})$ according to the continuum equation
\[
\frac{\partial\hat{\bf L}_d({\bf r},t)}{\partial t}=\frac{1}{i\hbar}\left[\hat{\bf L }_d({\bf r},t),\hat{H}_0\right]=-\nabla\cdot\hat{{\bf J}}_d({\bf r})-\hat{\bf T}_d({\bf r}).
\]
The OAM current density operator 
    \begin{align}
     \hat{\bf J}_{d}({\bf r})
     &=\frac{1}{V}\sum_{\alpha,\beta}\sum_{{\bf k},{\bf k'}}\left(\frac{\hbar^2 {\bf k'}}{2m^h_\beta}+\frac{\hbar^2 {\bf k}}{2m^h_\alpha}\right)\otimes\tilde{\pmb{\cal L}}^h_{\bf kk'}|_{\alpha\beta}\nonumber\\
     &\times \hat{h}^\dagger_{\alpha,{\bf k}}\hat{h}_{\beta,{\bf k'}}e^{i({\bf k'}-{\bf k})\cdot{\bf r}}.
     \label{OAM_current_density}
\end{align}
It can be alternatively derived according to the anticommutator of the velocity $\hat{\bf v}$ and OAM density $\hat{\bf L}_d({\bf r})$ operators $\hat{J}^{i}_{d,j}({\bf r})=\{\hat{v}_i,\hat{ L}_{d,j}({\bf r})\}/2$, in which the superscript $``i"$ defines the current flow direction and the subscript $``j"$ denotes the polarization direction of the OAM. This definition is consistent with the total OAM current $\hat{J}^i_{j}=\int d{\bf r} \hat{J}^{i}_{d,j}({\bf r})=\{\hat{v}_i,\hat{ L}_j\}/2$, defined as the spatial integral over the OAM current density~\cite{go2018intrinsic,Bhowal1,Bhowal2,Vignale1,Pezo,Pezo2023,Culcer}.
The OAM torque density operator (see Supplementary Material for the derivation of orbital current, orbital torque, and orbital injection rate from the Luttinger Hamiltonian)
\begin{align}
  \hat{\bf T}_d({\bf r})&=\frac{1}{V}\sum_{\alpha,\beta}\sum_{{\bf k},{\bf k'}}\left(\frac{i\hbar^2}{2m^h_\beta}-\frac{i\hbar^2}{2m^h_\alpha}\right)\left({\bf k'}\cdot{\bf k}\right)\tilde{\boldsymbol{\cal L}}^h_{\bf kk'}|_{\alpha\beta}\nonumber\\
  &\times \hat{h}^\dagger_{\alpha,{\bf k}}\hat{h}_{\beta,{\bf k'}}e^{i({\bf k'}-{\bf k})\cdot{\bf r}},
  \label{OAM_torque_density}
\end{align}
which originates from the inter-band elements in the OAM density operator~\eqref{OAMD_operator}.
The orbital torque to OAM vanishes when $m^h_\alpha=m^h_\beta$, implying that the OAM is conserved when the crystal field is absent such that the atomic $p$-orbitals remain degenerate. 
The total orbital torque operator $\int d{\bf r}\hat{T}_{d,j}({\bf r}) =[\hat{H}_0,\hat{ L}_j]/(2i)$ is a spatial integration over the OAM torque density, which is expressed as the commutator between the Hamiltonian $\hat{H}_0$ and the OAM  operator $\hat{\bf L}$.  This implies that the OAM is generally not conserved in the crystal since the periodic crystal potential breaks the $SO(3)$ rotation symmetry.

\textit{Evanescent Orbital Pumping}.---We sketch the principle of orbital pumping driven by an electromagnetic field. The effect is ``evanescent" since the stray field decays exponentially away from the source in a long range that renders the magnetic nanostructures not necessarily in contact with the metals or semiconductors, as illustrated in Fig.~\ref{model}.
We show that the photonic spin of the electromagnetic field can be converted to the OAM by defining the OAM injection rate. 
With the density matrix operator $\hat{\rho}^I$ in the interaction representation $``I"$, the rate of change of the OAM density
\begin{align}
    \dot{\bf L}_d({\bf r})={\rm Tr}(\hat{\rho}^I(t)\dot{\hat{{\bf L}}}_d^I({\bf r},t))+{\rm Tr}(\dot{\hat{\rho}}^I(t)\hat{\bf L}_d^I({\bf r},t)),
\end{align}
in which on the right-hand side the first term leads to $-\nabla\cdot\langle\hat{\bf J}_d({\bf r})\rangle-\langle\hat{\bf T}_d({\bf r})\rangle$ according to the continuum equation, while the second term describes the OAM injection rate density ${\bf R}_d({\bf r})$ into electron gas from the AC electromagnetic field.

Substituting the Liouville equation  $i\hbar\dot{\hat{\rho}}^I=[\hat{H}_{\rm int},\hat{\rho}^I]$, the DC orbital injection rate density reads (refer to the Supplementary Material for details) 
\begin{align}
 &{\bf R}^{\rm DC}_d({\bf r})={\rm Tr}\left(\dot{\hat{\rho}}^I\hat{\bf L}^I_d({\bf r})\right)_{\rm DC}=\left\langle\frac{1}{i\hbar}\left[\hat{\bf L}_d({\bf r}),\hat{H}_{\rm int}\right]\right\rangle_{\rm DC}\nonumber\\
  &=\frac{\mu_0\mu_B}{\hbar}\sum_{\zeta=\pm}\frac{m_0D}{\hbar^2}{\bf H}^{(\zeta)}({\bf r})\times\left\langle\hat{\bf L}_d({\bf r})\right\rangle^{(-\zeta)}\nonumber\\
  &+\sum_{\zeta=\pm}\sum_{a,b,c,j}\frac{\zeta e}{i\hbar\omega }\chi_{abc}^j\hat{\bf e}_a\{E_b^{(\zeta)}({\bf r}),\nabla_c\}\left\langle\hat{L}_{d,j}({\bf r})\right\rangle^{(-\zeta)},
\end{align}
where the coefficient tensor 
$\chi_{abc}^a=-r\delta_{ab}\delta_{bc}+(2A-r)\delta_{bc}$, $\chi_{abc}^b=-(r/2)\delta_{ac}(1-\delta_{bc})$, and $\chi_{abc}^c=-(r/2)\delta_{ab}(1-\delta_{bc})$.
We thereby interpret the orbital injection rate as the external torque due to the AC field on the OAM.  The magnetic field induced OAM density $\langle\hat{\bf L}_d({\bf r})\rangle^{(-\zeta)}\propto {\bf H}^{(\zeta)*}$ in the linear-response regime, so the DC component of the OAM injection rate $\propto {\bf H}^{(\zeta)}\times{\bf H}^{(\zeta)*}$ may be interpreted as the conversion of the photonic spin into the OAM of electrons~\cite{nukui2025light,zhang2008laser,noncontact_spin_pumping}. The local electric field induces an inhomogeneous OAM in the linear response, which is further driven, leading to the other contribution to the DC orbital injection.
In the steady state, $\dot{\bf L}=0$, so a balance occurs among the orbital-current, orbital-torque, and orbital-injection-rate densities ${\bf R}_d({\bf r})=\nabla\cdot\langle\hat{\bf J}_d({\bf r})\rangle+\langle\hat{\bf T}_d({\bf r})\rangle$, indicating that the OAM injected from the AC field is converted into the OAM current and intrinsic OAM torque.

We proceed to derive the spatial distribution of the OAM current and the intrinsic OAM torque.
The elements of the density matrix $\rho_{\bf kk'}|_{\alpha\beta}=\langle\alpha,{\bf k}|\hat{\rho}|\beta,{\bf k}'\rangle$ under the basis of the Bloch states $|\alpha,{\bf k}\rangle=\hat{h}^\dagger_{\alpha,{\bf k}}|0\rangle$ in the Schr\"{o}dinger picture obey the Liouville equation
\begin{align}
    &i\hbar\frac{\partial \rho_{\bf kk'}|_{\alpha\beta}}{\partial t}=\left(\varepsilon^h_{\alpha,{\bf k}}-\varepsilon^h_{\beta,{\bf k}'}\right)\rho_{\bf kk'}|_{\alpha\beta}\nonumber\\&+\sum_{\zeta=\pm}\sum_{\gamma,{\bf q}}\left(G^{h,(\zeta)}_{\bf kq}|_{\alpha\gamma}\rho_{\bf qk'}|_{\gamma\beta}-\rho_{\bf kq}|_{\alpha\gamma}G^{h,(\zeta)}_{\bf qk'}|_{\gamma\beta}\right)e^{-i\zeta\omega t}.
    \nonumber
\end{align}
By introducing the interaction adiabatically by $\hat{H}_{\rm int}(t)\rightarrow\hat{H}_{\rm int}(t)e^{\varrho t/\hbar}|_{\varrho\rightarrow0_+}$, we find the perturbative solution up to the second order of interaction~\cite{LandauLifshitz}
\begin{align}
   &\rho_{\bf kk'}|_{\alpha\beta}\approx
   \delta_{\bf kk'}\delta_{\alpha\beta}f(\varepsilon^h_{\alpha,{\bf k}})\nonumber\\
   &+\sum_{\zeta=\pm}\frac{f(\varepsilon^h_{\beta,{\bf k'}})-f(\varepsilon^h_{\alpha,{\bf k}})}{\varepsilon^h_{\beta,{\bf k'}}+\zeta\hbar\omega-\varepsilon^h_{\alpha,{\bf k}}+i\varrho}G^{h,(\zeta)}_{\bf kk'}|_{\alpha\beta}e^{-i\zeta\omega t}\nonumber\\
&+\sum_{\zeta_1,\zeta_2=\pm}\sum_{\gamma,{\bf q}}\frac{G^{h,(\zeta_1)}_{\bf kq}|_{\alpha\gamma}G^{h,(\zeta_2)}_{\bf qk'}|_{\gamma\beta}e^{-i(\zeta_1+\zeta_2)\omega t}}{\varepsilon^h_{\beta,{\bf k'}}+(\zeta_1+\zeta_2)\hbar\omega-\varepsilon^h_{\alpha,{\bf k}}+2i\varrho}\nonumber\\
&\times\left(\frac{f(\varepsilon^h_{\beta,{\bf k'}})-f(\varepsilon^h_{\gamma,{\bf q}})}{\varepsilon^h_{\beta,{\bf k'}}+\zeta_2\hbar\omega-\varepsilon^h_{\gamma,{\bf q}}+i\varrho}+\frac{f(\varepsilon^h_{\alpha,{\bf k}})-f(\varepsilon^h_{\gamma,{\bf q}})}{\varepsilon^h_{\gamma,{\bf q}}+\zeta_1\hbar\omega-\varepsilon^h_{\alpha,{\bf k}}+i\varrho}\right),
\nonumber
\end{align}
where $f(\varepsilon^h_{\alpha,{\bf k}})=1/(e^{(\varepsilon
^h_{\alpha,\mathbf{k}}-\mu)/(k_{B}T)}+1)$ is the
Fermi-Dirac distribution at temperature $T$. The inhomogeneous AC field establishes correlations between the Bloch states of different wave vectors and bands. The linear order of the electromagnetic field contributes to an AC response, while its quadratic order contributes to both the AC ($\zeta_1=\zeta_2$) and DC ($\zeta_1=-\zeta_2$) responses.

According to ${\bf J}_d({\bf r})={\rm Tr}({\hat{\rho}\hat{\bf J}_d({\bf r})})$, we obtain the pumped DC current density of OAM
 \begin{align}
    &{\bf J}^{\rm DC}_d({\bf r})
   \approx\frac{\hbar}{2V}\sum_{\zeta=\pm}\sum_{\alpha,\beta,\gamma}\sum_{\bf k,k',q}\left(f(\varepsilon^h_{\gamma,{\bf q}})-f(\varepsilon^h_{\alpha,{\bf k}})\right)\nonumber\\
    &\times\frac{G^{h,(\zeta)}_{\bf kq}|_{\alpha\gamma}G^{h,(-\zeta)}_{\bf qk'}|_{\gamma\beta}}{(\varepsilon^h_{\alpha,{\bf k}}-\zeta\hbar\omega-\varepsilon^h_{\gamma,{\bf q}}-i\varrho)(\varepsilon^h_{\beta,{\bf k'}}-\zeta\hbar\omega-\varepsilon^h_{\gamma,{\bf q}}+i\varrho)}\nonumber\\&\times\left(\frac{\hbar{\bf k'}}{m^h_\beta}+\frac{\hbar{\bf k}}{m^h_\alpha}\right)\otimes\tilde{\boldsymbol{\cal L}}^h_{\bf k'k}|_{\beta\alpha}e^{i({\bf k}-{\bf k'})\cdot {\bf r}}.
    \label{orbital_current_DC}
\end{align}
The pumped OAM current depends linearly on photon intensity and frequency $\omega$, as detailed in the Supplementary Material.
On the other hand, the OAM torque density by ${\bf T}_d({\bf r})={\rm Tr}(\hat{\rho}\hat{\bf T}_d({\bf r}))$ reads 
\begin{align}
    &{\bf T}^{\rm DC}_d({\bf r})
   \approx i\frac{\hbar^3}{2V}\sum_{\zeta=\pm}\sum_{\alpha,\beta,\gamma}\sum_{\bf k,k',q}\left(f(\varepsilon^h_{\gamma,{\bf q}})-f(\varepsilon^h_{\alpha,{\bf k}})\right)\nonumber\\
    &\times\frac{G^{h,(\zeta)}_{\bf kq}|_{\alpha\gamma}G^{h,(-\zeta)}_{\bf qk'}|_{\gamma\beta}}{(\varepsilon^h_{\alpha,{\bf k}}-\zeta\hbar\omega-\varepsilon^h_{\gamma,{\bf q}}-i\varrho)(\varepsilon^h_{\beta,{\bf k'}}-\zeta\hbar\omega-\varepsilon^h_{\gamma,{\bf q}}+i\varrho)}\nonumber\\
    &\times\left(\frac{1}{m^h_\beta}-\frac{1}{m^h_\alpha}\right)({\bf k}\cdot{\bf k}')\tilde{\boldsymbol{\cal L}}^h_{\bf k'k}|_{\beta\alpha}e^{i({\bf k}-{\bf k'})\cdot {\bf r}}.
    \label{orbital_Torque_DC}
\end{align}

\textit{Orbital current $vs.$ orbital torque in materials}.---We illustrate the evanescent orbital pumping by the electromagnetic field generated by the ferromagnetic resonance (FMR) of a magnetic nanowire of width $w$, thickness $d$, and saturation magnetization along the wire $\hat{\bf y}$-direction as in Fig.~\ref{model}.
With the amplitude of the excited transverse magnetization $M_z$, the radiated  AC magnetic field in the wave-vector space~\cite{Jackson} (see the detailed derivation in the Supplementary Material)
\begin{align}
    \left(\begin{array}{cc}H^{(\zeta)}_z({\bf k})\\
    H^{(\zeta)}_x({\bf k})\end{array}\right)&=-\zeta\frac{i}{2} \frac{1-e^{-\left|k_x\right| d}}{|k_x|-ik_z} \frac{ \sin \left(k_x w / 2\right)}{k_x}\delta(k_y)\nonumber\\
    &\times\left(1-\zeta{\rm sgn}(k_x) \right)\left(\begin{array}{cc}1\\i{\rm sgn}(k_x)\end{array}\right)M_z
    \label{Fourier_FMR}
\end{align}
is circularly polarized around the wire $\hat{\bf y}$-direction when $w=d$, with circular polarization locked to the wave vector $k_x$; the electric field
\begin{align}
E_y^{(\zeta)}({\bf k})=({\zeta\mu_0\omega}/{k_x})H_z^{(\zeta)}(k_x,z)
\label{electric_field}
\end{align}
is linearly polarized along the wire $\hat{\bf y}$-direction.

\begin{figure}[pth]
\centering
\hspace{0cm}\includegraphics[width=0.46\textwidth, trim=0cm 0.1cm 0.0cm 0cm, clip]{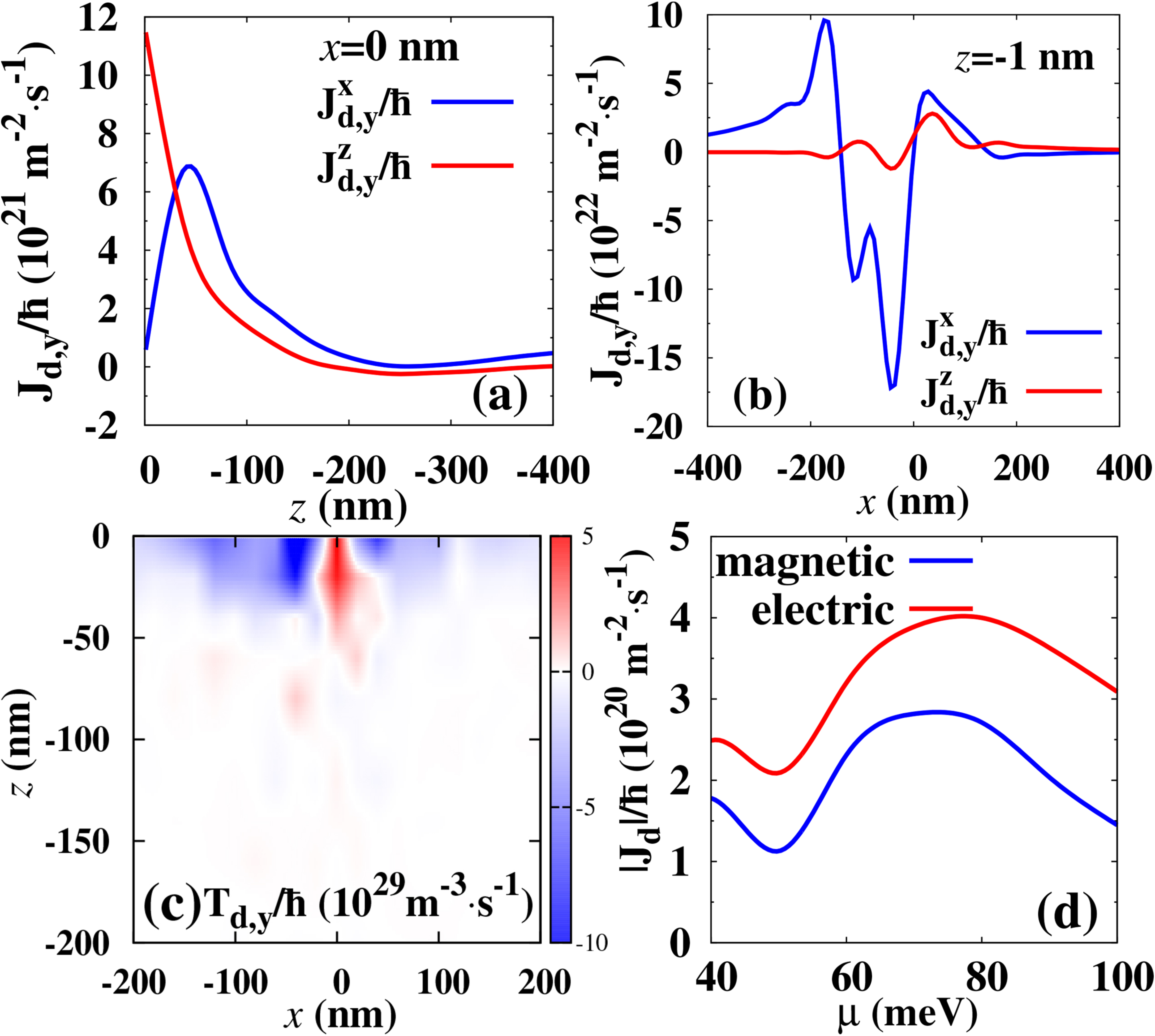}
\caption{Spatial distribution of longitudinal OAM current density of holes when pumped by the electromagnet field emitted by the FMR of a magnetic nanowire. (a) and (b) compare the flow along the $-\hat{\bf z}$- and $\hat{\bf x}$-directions at $x=0$~nm and $z=-1$~nm, respectively. (c) plots the orbital torque density polarized along the magnetization $\hat{\bf y}$-direction. (d) compares the contribution of electric and magnetic fields in the orbital pumping with different chemical potentials at position $(x,z)=(0,-200)$~nm.}
\label{orbital_current}
\end{figure}

\begin{figure}[pth]
\centering
\includegraphics[width=0.48\textwidth, trim=0.1cm 0cm 0.0cm 0.0cm, clip]{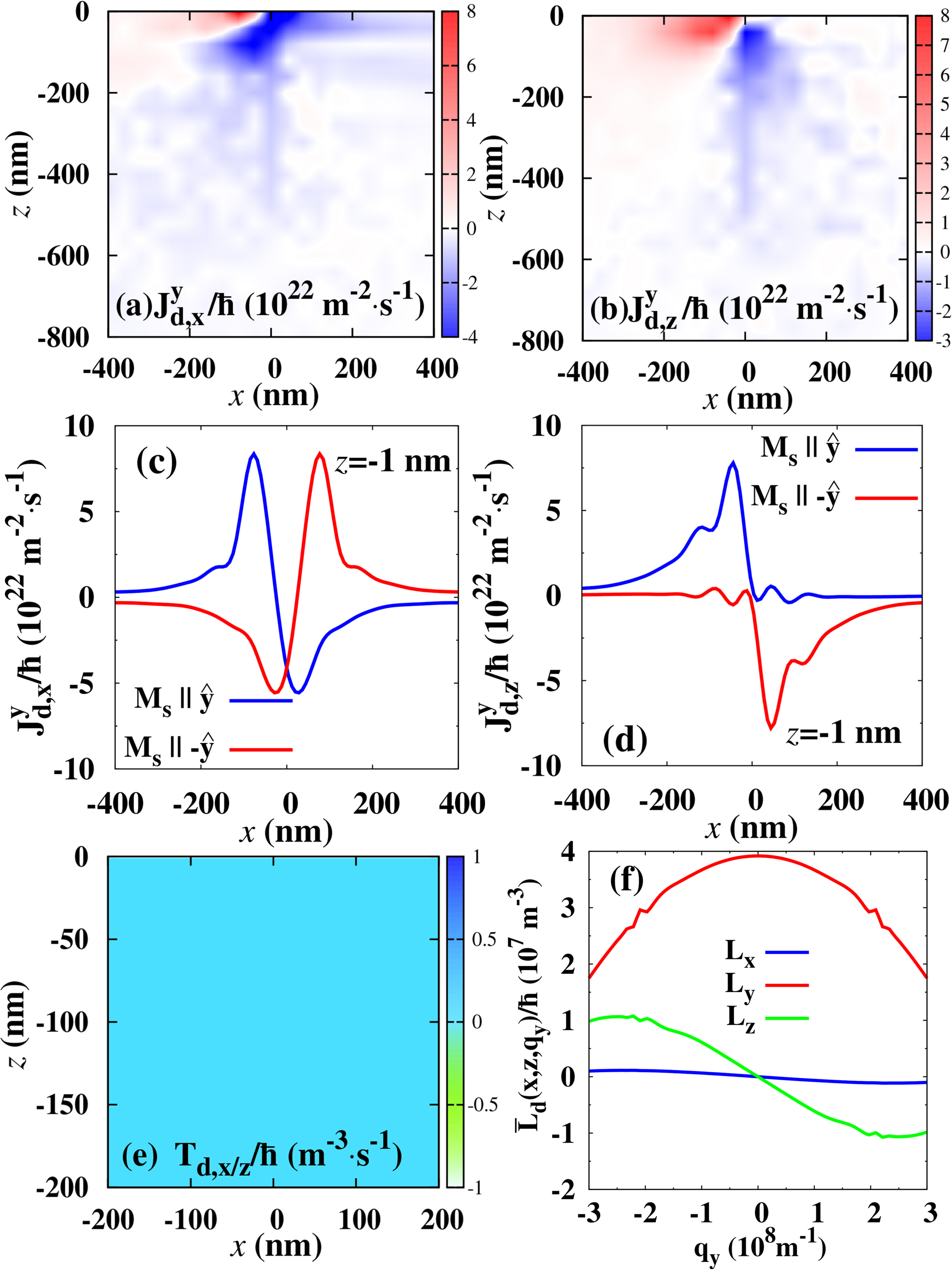}
\caption{Spatial distribution of orbital Hall current density flowing along the wire $\hat{\bf y}$-direction with orbital polarization along $\hat{\bf x}$ (a) and $\hat{\bf z}$ (b). (c) and (d) compare the orbital Hall current polarized along $\hat{\bf x}$ and $\hat{\bf z}$ at two sides of the nanomagnet. (e) plots the orbital torque density polarized normal to the magnetization $\hat{\bf y}$-direction. (f) Calculated DC OAM density $\overline{\bf L}_d(x,z,q_y)$ in the mixed spatial $(x,z)$ and wave-vector $q_y$ space at position $(x=0~{\rm nm},z=-100~{\rm nm})$.}
\label{OAM torque}
\end{figure}

Figure~\ref{orbital_current} illustrates the spatial distribution of the longitudinal OAM current density along the gradient of the field, calculated using the typical semiconductor material parameters: $A=-2\times10^{-37}$~${\rm J}\cdot{\rm m}^2$, $r=-1.5\times10^{-37}$~${\rm J}\cdot{\rm m}^2$, and $D=-1.5\times10^{-37}$~${\rm J}\cdot{\rm m}^2$~\cite{Luttinger1956,Lawaetz}.
We note again that in the orbital current ${\bf J}^i_{d,j}$ the superscript $``i"$ defines the current flow direction and the subscript $``j"$ denotes the polarization direction of the OAM. We choose the chemical potential of holes $\mu=60$~meV.
The AC stray field is generated by CoFeB of $d=w=50$~nm and saturation magnetization $\mu_0M_s=1.6$~T, biased by a static magnetic field $\mu_0H_0=0.1$~T along the wire $\hat{\bf y}$-direction. A small transverse magnetization  $M_z=0.05M_s$ is excited in the FMR at a frequency $2\pi\times 26$~GHz by uniform microwaves. 
In Fig.~\ref{orbital_current}(a) and (b),  the DC OAM current polarized along the photonic spin $\hat{\bf y}$-direction flows \textit{asymmetrically} outward from the source. The transport of the OAM is anisotropic: the DC OAM current decays much more slowly along the surface normal $-\hat{\bf z}$-direction than that along the surface $\hat{\bf x}$-direction. Holes with orbital polarization along the photonic spin/magnetization $\hat{\bf y}$-direction experience the OAM torque density $T_{d,y}\hat{\bf y}$ as in Fig.~\ref{OAM torque}(c) for its spatial distribution, which concentrates on the region of near AC  fields and suggests that the OAM along $\hat{\bf y}$ is not conserved. The electric field contributes dominantly when the chemical potential is large, as in Fig.~\ref{OAM torque}(d).

Unexpectedly, evanescent orbital pumping also generates a transverse/Hall OAM current that is normal to the field gradient, as illustrated by the OAM current along the magnetic wire in Fig.~\ref{model}. Figure~\ref{OAM torque}(a) and (b) plot the orbital Hall current density carried by holes flowing along the wire $\hat{\bf y}$-direction with orbital polarization normal to the wire, i.e., along the $\hat{\bf x}$- and $\hat{\bf z}$-directions, respectively. This can be intuitively interpreted by the joint effect of magnetic and electric fields: the magnetic field \eqref{Fourier_FMR}, normal to the wire, polarizes the OAM, while the electric field \eqref{electric_field} drives the current along the wire. 
The orbital Hall current is not symmetric at the two sides of the wire, which is attributed to the chirality of the near electromagnetic field. Indeed, when the chirality is reversed with the saturation magnetization switched to ${\bf M}_s\parallel -\hat{\bf y}$, the asymmetric distribution of the orbital Hall current is switched, as shown in Fig.~\ref{OAM torque}(c) and (d). 
In our perturbation theory, the OAM torque $T_{d,x}\hat{\bf x}$ and $T_{d,z}\hat{\bf z}$ vanish, as in Fig.~\ref{OAM torque}(e), indicating the transverse OAM Hall current is intrinsically conserved, a merit attractive for orbital transport.

The mechanism for generating the orbital Hall current is as follows. When the system holds a translational symmetry along the $\hat{\bf y}$-direction, the pumped DC OAM density ${\bf L}^{\rm DC}_d({\bf r})=\sum_{q_y}\overline{\bf L}_d(x,z,q_y)$ is independent of $y$, in which  $\overline{\bf L}_d(x,z,q_y)$ may be interpreted as the DC OAM density at position $(x,z)$ carried by holes of wave vector $q_y$. 
We demonstrate from Eq.~(\ref{OAMD_operator}) that under an operation $q_y\rightarrow -q_y$, the components $\overline{L}_{d,y}(x,z,q_y)=\overline{L}_{d,y}(x,z,-q_y)$ and $\overline{L}_{d,x/z}(x,z,q_y)=-\overline{L}_{d,x/z}(x,z,-q_y)$, as shown in Fig.~\ref{OAM torque}(f). Therefore, the excited hole OAM propagating along the $+\hat{\bf y}$- and $-\hat{\bf y}$-directions carry the same $\hat{\bf y}$-polarization, but the opposite $\hat{\bf x}/\hat{\bf z}$-polarization.
The latter contributes to an orbital Hall current ${J}_{d,x/z}^{y}|_{\rm DC}$ flowing along the wire $\hat{\bf y}$-direction with the orbital polarization along $\hat{\bf x}/\hat{\bf z}$, as in Figs.~\ref{OAM torque}(a) and (b).

\textit{Conclusion and discussion}.---In conclusion, we have predicted a different mechanism of orbital pumping that does not rely on the SOC via the driven dynamics of the orbital magnetic moments by local electromagnetic fields applied to the semiconductors/metals. According to this mechanism, the focused laser beams may also act as an efficient source for optical orbital pumping. 
The orbital torque of the AC field to the OAM causes both longitudinal and transverse/Hall orbital currents, which flow along and normal to the field gradient. The latter flow is immune to the orbital torque and is expected to propagate long distances. The proposed effect is evanescent, avoiding the interfacial exchange proximity effect, and is efficient for semiconductors such as Si and Ge.  Experimental detection of the predicted OAM currents could be feasible using magneto-optical or nitrogen-vacancy center imaging techniques or indirect probes like orbital-induced magnetoresistance effects. Our study adds a fundamental understanding of the OAM dynamics and offers guidance for designing semiconductor-based orbitronic devices operating without significant dissipation.

\begin{acknowledgments}
This work is financially supported by the National Key Research and Development Program of China under Grant No.~2023YFA1406600 and the National Natural Science Foundation of China under Grant No.~12374109. H.W. acknowledges the support of Swiss
National Science Foundation (Grant No. 200020-200465) and the China Scholarship Council (CSC, Grant No. 202206020091). 
\end{acknowledgments}

\end{document}